\documentclass{aa}  
\usepackage{natbib}
\bibpunct{(}{)}{;}{a}{}{,} % to follow the A&A style

\usepackage{graphicx}
\usepackage{txfonts}
\newcommand{\va}{v_{\mathrm{A}}}
\newcommand{\vs}{v_{\mathrm{s}}}

\begin{document}

	\title{Stability of thermal modes in cool prominence plasmas}

	\titlerunning{Thermal modes in prominences}

   \author{R. Soler\inst{\ref{leuven}} \and J. L. Ballester\inst{\ref{uib}} \and S. Parenti\inst{\ref{rob}}}
\offprints{R. Soler}
\institute{Centre for Plasma Astrophysics, Department of Mathematics, KU Leuven,
              Celestijnenlaan 200B, 3001 Leuven, Belgium.  \\ \email{roberto.soler@wis.kuleuven.be}
 \label{leuven}
 \and
 Solar Physics Group, Departament de F\'isica, Universitat de les Illes Balears,
             E-07122 Palma de Mallorca, Spain. \\ \email{joseluis.ballester@uib.es} \label{uib} 
\and
Royal Observatory of Belgium, 3 Av. Circulaire, 1180 Bruxelles, Belgium. \\ \email{s.parenti@oma.be} \label{rob}             
             }

 	 \date{Received XXX / Accepted XXX}

  \abstract
   {Magnetohydrodynamic thermal modes may play an important role in the formation, plasma condensation, and evolution of solar prominences. Unstable thermal modes  due to  unbalance between radiative losses and heating can lead to rapid plasma cooling and condensation. An accurate description of the radiative loss function is therefore crucial for this process.}
   {We study the stability of thermal modes in unbounded and uniform plasmas with properties akin to those in solar prominences. Effects due to partial ionization are taken into account. Three different parametrizations of the radiative loss function are used.}
   {By means of a normal mode analysis, we investigate  linear nonadiabatic  perturbations superimposed on the equilibrium state. We find an approximate instability criterion for thermal modes, while the exact linear growth rate is obtained by numerically solving the general dispersion relation. The stability of thermal disturbances   is compared for the three different  loss functions considered.}
   {Using up-to-date computations of radiative losses derived from the CHIANTI atomic database, we find that thermal modes may be unstable in prominences for lower temperatures than those  predicted with previously existing loss functions. Thermal instability can take place for temperatures as low as 15,000~K, approximately.  The obtained linear growth rates indicate that this instability might have an important impact on the dynamics and evolution of cool prominence condensations.}
   {}


   \maketitle

%________________________________________________________________

\section{Introduction}

Thermal or condensational modes have been extensively investigated in magnetized plasmas \citep[e.g.,][]{parker,field,heyvaerts}. As explained by \citet{parker}, thermal instability can happen in a diffuse medium due to unbalance between temperature-independent energy gains, i.e., heating, and temperature-dependent radiative losses. \citet{parker} arrived at the qualitative criterion that instability can be present when radiative losses decrease as the temperature increases. This mechanism is important in the context of prominences since it allows the formation of cool plasma condensations in a medium of high temperature. Therefore, unstable thermal modes may play an important role in the formation of solar prominences and in the evolution of the prominence plasma. \citet{field} and \citet{heyvaerts} investigated the phenomenon in more detail than \citet{parker} and derived more accurate instability criteria. Subsequent papers that investigate both linear and nonlinear thermal instabilities, mainly in the field of prominences, are, e.g., \citet{hildner,oran,dahlburgmariska,karpen,cargillhood,vanderlinden91,carbonell04,solercont} among others.

As shown in the instability criterion derived by \citet{field}, an accurate description of the radiative loss function is crucial to ascertain the stability of thermal modes. However, the determination of the radiative loss function in prominence plasmas depending on the values of  temperature and density is a  difficult work that requires complicated  numerical solutions of the  radiative transfer equations for nonlocal thermodynamic equilibrium.  Alternatively, several semi-empirical parametrizations of the radiative loss function for prominence and coronal conditions are available in the literature. These parametrizations enable us to incorporate radiative losses in a consistent way in the theoretical models of prominence plasmas  without the need of  solving the  radiative transfer problem. 

 One of the first parametrizations of this kind can be found in \citet{hildner}, who performed a piecewise fit as function of the temperature of the computations of radiative losses available by then. Subsequent authors have proposed different parametrizations that update Hildner's fit \citep[e.g.,][]{rosner,milne}, although Hildner's function is still used in some works nowadays. Another  function frequently used in the literature is the so-called Klimchuk-Raymond function \citep[see, e.g.,][]{klimchuk}  that is a better representation of the radiative losses for  prominence-corona transition region and coronal temperatures. The shape of the loss function depends on the completeness of the atomic model used for the calculation, on the atomic processes included, on the ionization equilibrium, and element abundance assumed.  More recent loss functions that incorporate accurate atomic data information are, e.g., the loss function used by \citet{parenti2006} and \citet{parentivial} computed from the CHIANTI database \citep{dere1997,landi}, and the loss function computed by \citet{schure} using the SPEX package \citep{kaastra}. 

The purpose of this paper is to investigate the stability properties of thermal modes in cool prominence plasmas. We  compare the results using up-to-date computations of radiative losses derived from the CHIANTI v7 atomic database with those obtained assuming two of the most used loss functions existing in the literature. These radiative losses are obtained assuming an optically thin plasma, while the core of prominence cannot completely satisfy this condition. We will discuss our assumption later on. This paper is organized as follows. Section~\ref{sec:basic} contains a description of the equilibrium and the basic equations.  The instability criterion for thermal modes is derived in Section~\ref{sec:crit}. Parametric studies of the linear growth rate are done in Section~\ref{sec:res}. Finally, Section~\ref{sec:discussion} contains the summary and discussion of the results.

\section{Basic equations}
\label{sec:basic}

\subsection{Equilibrium}

Our equilibrium configuration is a uniform plasma of infinite extend. We assume a partially ionized hydrogen plasma composed of ions, electrons, and neutrals. We use Cartesian coordinates and all quantities are expressed in MKS units thorough this paper. The magnetic field, $\bf B$, is uniform and orientated along the $z$-direction, namely ${\bf B} = B \hat{e}_z$, with $B$ constant. We denote by $\rho$, $T$, and $p$ the equilibrium mass density, temperature, and gas pressure, respectively.  The set of basic nonadiabatic MHD equations governing the plasma dynamics in the single-fluid approximation are \citep[see, e.g.,][]{brag}
\begin{eqnarray}
\frac{{\rm D}\rho}{{\rm D} t} &=&- \rho \nabla \cdot {\bf v}, \label{eq:cont} \\
\rho \frac{{\rm D} {\bf v}}{{\rm D} t} &=& - \nabla p + \frac{1}{\mu} \left( \nabla \times {\bf B} \right) \times {\bf B}, \label{eq:mom} \\
\frac{\partial {\bf B}}{\partial t} &=& \nabla \times \left( {\bf v} \times {\bf B} \right) - \nabla \times \left(  \eta \nabla \times {\bf B} \right) \nonumber \\ &+& \nabla \times \left\{ \frac{\eta_{\rm C} - \eta}{{\bf B}^2} \left[ \left( \nabla \times {\bf B} \right) \times {\bf B} \right] \times {\bf B}   \right\}, \label{eq:induct} \\
\frac{{\rm D} p}{{\rm D} t} &=& \frac{\gamma p}{\rho} \frac{{\rm D} \rho}{{\rm D} t} + \left( \gamma -1  \right) \left[ \nabla \cdot \left( \kappa \nabla T \right) - \rho L \left(  \rho, T\right) \right],\label{eq:energy} \\
p&=&\frac{\rho R T}{\tilde{\mu}}, \label{eq:gaslaw}
\end{eqnarray}
where $\frac{{\rm D}}{{\rm D} t} = \frac{\partial}{\partial t} + {\bf v} \cdot \nabla$ is the material derivative for time variations following the plasma motion, $\bf v$ is the plasma velocity, $\mu$ is the magnetic permittivity, $\gamma$ is the adiabatic index, $\kappa$ is the thermal conductivity tensor, $L \left(  \rho, T\right)$ is the heat-loss function, $\eta$ and $\eta_{\rm C}$ are the coefficients of Ohm's and Cowling's diffusion, respectively, $R$ is the ideal gas constant, and $\tilde{\mu} = \frac{1}{1+\xi_{\rm i}}$ is the mean atomic weight, with $\xi_{\rm i}$ the relative fraction of ions. This parameter ranges from $\xi_{\rm i}=1$ in a fully ionized plasma and $\xi_{\rm i}=0$ in a neutral gas. In Equations~(\ref{eq:cont})--(\ref{eq:gaslaw}), the effects of gravity and viscosity have been omitted. Equation~(\ref{eq:induct}) is  the induction equation.  In Equation~(\ref{eq:induct}) we have neglected some minor terms which are several orders of magnitude smaller than Ohm's and Cowling's terms in partially ionized prominence plasmas \citep[see an expression of the complete single-fluid induction equation in, e.g.,][]{forteza,temury}. The neglected terms are Hall's term, the diamagnetic current term, and Biermann's battery term. \citet{soler09res} showed that the effect of Hall's term on the waves is negligible in prominence conditions. The diamagnetic current term and  Biermann's battery term are only relevant when large pressure gradients are present, a situation more representative of stellar interiors. The omission of these terms in the present work is therefore justified.  In a partially ionized plasma, Cowling's diffusion represents an enhanced magnetic diffusion due to ion-neutral collisions. The expressions for $\eta$ and $\eta_{\rm C}$ are
\begin{eqnarray}
\eta &=& 3.7 \times 10^{-6} \frac{m_{\rm e} \ln \Lambda_{\rm C}}{\mu e^2} T^{-3/2}, \\
\eta_{\rm C} &=& \eta + \frac{B^2 \left( 1 - \xi_{\rm i} \right)^2}{\mu \alpha_{\rm n}},
\end{eqnarray}
where $m_{\rm e}$ is the electron mass, $e$ is the electron charge, $\ln \Lambda_{\rm C}$ is Coulomb's logarithm \citep[see, e.g.,][]{priest}, and $\alpha_{\rm n}$ is the neutral friction coefficient given by
\begin{equation}
\alpha_{\rm n} = \frac{1}{2} \xi_{\rm i} \left( 1 - \xi_{\rm i} \right) \frac{\rho^2}{m_{\rm p}} \sqrt{\frac{16 k_{\rm B} T}{\pi m_{\rm p}}},
\end{equation}
with $m_{\rm p}$ the proton mass and $k_{\rm B}$ Boltzmann's constant. In the fully ionized case, $\eta_{\rm C} = \eta$ and the third term on the right-hand side of Equation~(\ref{eq:induct}) is absent.

We denote by $\kappa_\parallel$ and $\kappa_\perp$ the parallel and perpendicular scalar components of the thermal conductivity tensor with respect to the magnetic field direction, which can be expressed as
\begin{equation}
\kappa_\parallel = \kappa_{\rm e} + \kappa_{\rm n}, \qquad \kappa_\perp = \kappa_{\rm i} + \kappa_{\rm n} \approx \kappa_{\rm n}
\end{equation}
with $\kappa_{\rm e}$, $\kappa_{\rm i}$, and $\kappa_{\rm n}$ the conductivities by electrons, ions, and neutrals, respectively. In a fully ionized medium,  $\kappa_\parallel$ is governed by electrons, whereas $\kappa_\perp$ is caused by ions. In a partially ionized plasma, the contribution of neutrals, $\kappa_{\rm n}$, has to be added to both scalar conductivities because  thermal conduction by neutrals is isotropic. Since $\kappa_{\rm i} \ll \kappa_{\rm n}$, the conductivity by ions can be neglected in $\kappa_\perp$. We use \citep[see, e.g.,][]{solerhelium}
\begin{equation}
\kappa_{\rm e} = 10^{-11} \xi_{\rm i} T^{5/2}, \qquad \kappa_{\rm n} = 2.24\times 10^{-2} \left( 1 - \xi_{\rm i} \right) T^{1/2}.
\end{equation}

\begin{figure}[!t]
\centering
 \includegraphics[width=0.99\columnwidth]{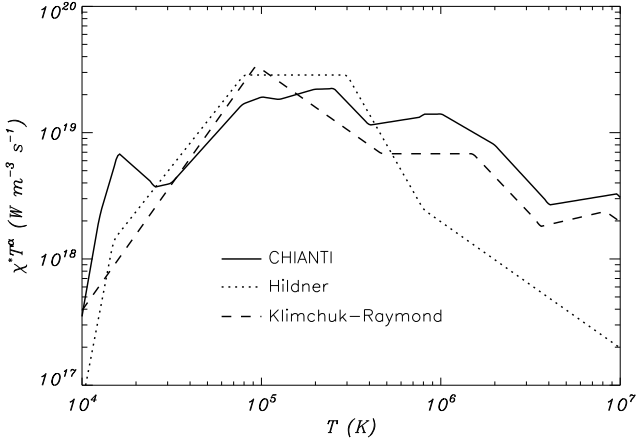}
\caption{Fit of the energy loss rate per unit volume as function of the temperature computed from the CHIANTI atomic database assuming coronal abundances, ionization equilibrium, and a constant pressure of 6.64~mPa (solid line) compared to Hildner's (dotted line) and Klimchuk-Raymond's fits (dashed line). The parameters $\chi^*$ and $\alpha$ of the three fits are given in Table~\ref{tab:parameters}. \label{fig:param}}
\end{figure}

Regarding the heat-loss function, $L \left(  \rho, T\right)$, we consider the following expression,
\begin{equation}
L \left(  \rho, T\right) = \rho \chi^* T^\alpha - h, \label{eq:heatloss}
\end{equation}
where $\chi^*$ and $\alpha$ are functions of the temperature, and $h$ is an arbitrary heating function. In the equilibrium we assume that radiative losses balance  heating, so  that $h$ is defined to satisfy $L \left(  \rho, T\right) =0$ in the equilibrium state. Several parametrizations of $\chi^*$ and $\alpha$ for prominence conditions are available in the literature. In this work, we use two of the most used loss functions, namely  the well-known parametrizations by Hildner \citep{hildner} and Klimchuk-Raymond \citep{klimchuk}. In addition, we use a parametrization of the radiative loss function computed from the CHIANTI v7 atomic database \citep{landi} assuming coronal abundances \citep{mazzotta}, ionization equilibrium, and a constant pressure of 6.64~mPa. The parameters $\chi^*$ and $\alpha$ corresponding to these three different fits are given in Table~\ref{tab:parameters}. Figure~\ref{fig:param} displays the loss rate per unit volume as function of the temperature for the three fits.  The main differences in the three curves are found at low ($T \lesssim 3\times 10^4$~K) and high ($T \gtrsim 10^6$~K) temperatures. For the region of our interest, that is the low temperature range, the peak at about $1.5 \times 10^4$~K in the CHIANTI calculation is due mainly to the H and He emissions. Besides this difference, it should be noted that the CHIANTI database is still incomplete at these temperatures, so that further increase of the loss function is expected for more complete calculations.

\begin{table}
\caption{Parameters of the loss function for the considered fits.}             % title of Table
\label{tab:parameters}      % is used to refer this table in the text
\centering                          % used for centering table
\begin{tabular}{l c c r}        % centered columns (4 columns)
\hline\hline                 % inserts double horizontal lines
Fit & Temperature Range (K) & $\chi^*$ & $\alpha$  \\    % table heading 
\hline                        % inserts single horizontal line
   CHIANTI & $T \leq 1.26\times 10^4$ & $2.02\times 10^{-15}$ & $8.06$  \\      % inserting body of the table
  & $1.26\times 10^4 < T \leq 1.58\times 10^4$ & $5.60\times 10^{-2}$ & $4.78$  \\
  & $1.58\times 10^4 < T \leq 2.51\times 10^4$ & $1.36\times 10^{24}$ & $-1.26$  \\
   & $2.51\times 10^4 < T \leq 3.16\times 10^4$ & $1.46\times 10^{17}$ & $0.32$  \\
      & $3.16\times 10^4 < T \leq 7.9\times 10^4$ & $3.11\times 10^{11}$ & $1.58$  \\
       & $7.9 \times 10^4 < T \leq 10^5$ & $4.44\times 10^{16}$ & $0.53$  \\
        & $10^5 < T \leq 1.25\times 10^5$ & $2.31\times 10^{20}$ & $-0.22$  \\
   & $1.25\times 10^5 < T \leq 2\times 10^5$ & $1.44\times 10^{17}$ & $0.41$  \\     
      & $2\times 10^5 < T \leq 2.51\times 10^5$ & $1.20\times 10^{19}$ & $0.05$  \\   
         & $2.51\times 10^5 < T \leq 3.98\times 10^5$ & $2.02\times 10^{27}$ & $-1.47$  \\     
         & $3.98\times 10^5 < T \leq 7.94\times 10^5$ & $6.38\times 10^{17}$ & $0.22$  \\     
           & $7.94\times 10^5 < T \leq 10^6$ & $1.40\times 10^{19}$ & $0.0$  \\     
           & $10^6 < T \leq 2\times 10^6$ & $1.26\times 10^{24}$ & $-0.82$  \\     
           & $2 \times 10^6 < T \leq 3.98\times 10^6$ & $4.14\times 10^{28}$ & $-1.54$  \\     
           & $3.98 \times 10^6 < T \leq 10^7$ & $7.74\times 10^{16}$ & $0.23$  \\     
           & $10^7 < T \leq 3.16 \times 10^7$ & $2.06\times 10^{25}$ & $-0.98$  \\     
           & $T > 3.16 \times  10^7 $ & $3.20\times 10^{16}$ & $0.20$  \\     
\hline         
Hildner & $T \leq 15\times 10^3$ & $1.76 \times 10^{-13}$ & $7.4$  \\      % inserting body of the table
   & $15\times 10^3 < T \leq 8\times 10^4$ & $4.29 \times 10^{10}$ &  $1.8$   \\
   & $8\times 10^4 < T \leq 3\times 10^5$& $2.86 \times 10^{19}$ &  $0.0$   \\
   & $3\times 10^5 < T \leq 8\times 10^5$ & $1.41 \times 10^{33}$ & $-2.5$  \\
   & $T > 8\times 10^5$ & $1.97 \times 10^{24}$ & $-1.0$  \\
\hline
Klimchuk- & $T \leq 9.33\times 10^{4}$ & $3.91 \times 10^{9}$ & $2.0$  \\
Raymond & $9.33\times 10^{4} < T \leq 4.68 \times 10^{5}$ & $3.18 \times 10^{24}$ & $-1.0$  \\
 & $4.68\times 10^{5} < T \leq 1.51 \times 10^{6}$ & $6.81 \times 10^{18}$ & $0.0$  \\
 & $1.51 \times 10^{6} < T \leq 3.55 \times 10^{6}$ & $1.27 \times 10^{28}$ & $-1.5$  \\
 & $3.55\times 10^{6} < T \leq 7.94 \times 10^{6}$ & $1.24 \times 10^{16}$ & $0.33$  \\
 & $ T > 7.94\times 10^{6}$ & $1.97 \times 10^{25}$ & $-1.0$  \\
\hline                          
\end{tabular}
\tablefoot{The fit derived from computations based on the CHIANTI atomic database assuming coronal abundances was used in \citet{parenti2006}. Hildner's and Klimchuk-Raymond's fits are adapted from \citet{hildner} and \citet{klimchuk}, respectively. Quantities are expressed in MKS units.} 
\end{table}

\subsection{Dispersion relation for linear perturbations}

We take the plasma initially at rest and superimpose linear perturbations on the equilibrium state. Equations~(\ref{eq:cont})--(\ref{eq:gaslaw}) are  linearized. We write perturbations proportional to $\exp \left( i {\bf k} \cdot {\bf r} - i  \omega t \right)$, where ${\bf r} = \left( x, y, z \right)$ is the position vector, ${\bf k} = \left( k_x, k_y, k_z \right)$ is the wavenumber vector, and $\omega$ is the frequency. For simplicity, we choose the reference frame so that we can set $k_y = 0$ and consider wave propagation in the $xz$-plane only. We focus our study on nonadiabatic magnetoacoustic and thermal modes. Alfv\'en modes are not discussed in the present investigation. We combine the linearized  Equations~(\ref{eq:cont})--(\ref{eq:gaslaw}) and, after some algebraic manipulations, we arrive at the dispersion relation for nonadiabatic magnetoacoustic and thermal modes \citep[see details in, e.g.,][]{carbonell04, solerhelium,barcelo}, which can be written in a compact form as follows,
\begin{equation}
\omega^4 - \left( \Gamma^2 + \Lambda^2 \right) k^2 \omega^2 +  k^4 \Lambda^2 \left[ \left( \Gamma^2 - \va^2 \right) + \va^2 \cos^2 \theta \right] = 0, \label{eq:reldisper}
\end{equation}  
where $k^2 = k_x^2 + k_z^2$ is the square of the wavevector modulus, $\cos \theta = k_z/k$ is the cosine of the angle between $\bf k$ and $\bf B$, $\Lambda^2 = \tilde{\gamma} p/\rho$ is the square of the nonadiabatic sound speed \citep[see][]{soler08}, with $\tilde{\gamma}$ the effective nonadiabatic index defined as
\begin{equation}
\tilde{\gamma} = \frac{\left( \gamma - 1 \right) \left[ \left( \tilde{\kappa}_\perp \sin^2 \theta + \tilde{\kappa}_\parallel \cos^2 \theta \right) k^2 + \omega_{\rm T} - \omega_\rho \right] - i \gamma \omega }{\left( \gamma - 1 \right) \left[ \left( \tilde{\kappa}_\perp \sin^2 \theta + \tilde{\kappa}_\parallel \cos^2 \theta \right) k^2 + \omega_{\rm T}\right] - i  \omega } ,
\end{equation}
with
\begin{equation}
\tilde{\kappa}_\parallel = \frac{T}{p} \kappa_\parallel, \qquad \tilde{\kappa}_\perp = \frac{T}{p} \kappa_\perp, \label{eq:defs1}
\end{equation}
\begin{equation}
\omega_{\rm \rho} = \frac{\rho^2}{p} \left( \frac{\partial L}{\partial \rho} \right)_T, \qquad \omega_{\rm T} = \frac{\rho T}{p} \left( \frac{\partial L}{\partial T} \right)_\rho. \label{eq:defs2}
\end{equation}
In addition, $\Gamma^2$ is the square of the modified Alfv\'en speed \citep[see][]{soler09} defined as
\begin{equation}
\Gamma^2 = \va^2 - i \omega \eta_{\rm C},
\end{equation}
with $\va^2 = B^2/\mu \rho$ the square of the ideal Alfv\'en speed.  Note that both $\Lambda^2$ and  $\Gamma^2$ are functions of  $\omega$ and $k$. In the absence of nonadiabatic effects, $\tilde{\kappa}_\parallel = \tilde{\kappa}_\perp = \omega_{\rm T} = \omega_\rho =  0$ and $\tilde{\gamma} = \gamma$, so that $\Lambda^2$ becomes the square of the adiabatic sound speed, $\vs^2 = \gamma p /\rho$. In the absence of magnetic diffusion, $\eta_{\rm C} = \eta = 0$ and $\Gamma^2$ becomes the square of the ideal Alfv\'en speed, $\va^2$. In such a case, Equation~(\ref{eq:reldisper}) reverts the the well-known dispersion relation for ideal, adiabatic magnetoacoustic waves in a plasma of infinite extend \citep[see, e.g.,][]{lighthill}.

For positive and real $k$ and $\theta$,  Equation~(\ref{eq:reldisper}) has five solutions of $\omega$. In general, $\omega$ is complex, namely $\omega=\omega_{\rm R}+i\omega_{\rm I}$, with $\omega_{\rm R}$ and $\omega_{\rm I}$ the real and imaginary parts of $\omega$, respectively. The solutions of Equation~(\ref{eq:reldisper}) were discussed in detail by \citet{carbonell04} and \citet{barcelo}. Of the five solutions of $\omega$, two complex conjugate solutions correspond to damped slow modes and other two complex conjugate solutions correspond to damped fast modes. The remaining solution is purely imaginary, i.e., $\omega_{\rm R} = 0$, and corresponds to the thermal mode. The thermal mode is the subject of our investigation.

\section{Approximate instability criterion}

\label{sec:crit}

\begin{figure}[!t]
\centering
 \includegraphics[width=0.95\columnwidth]{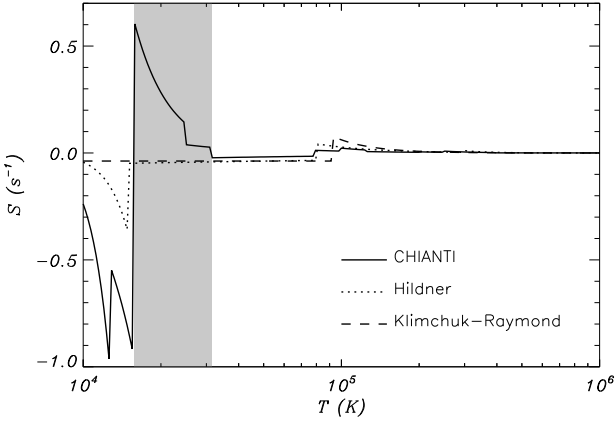}
\caption{Approximate thermal mode growth rate vs. temperature in the absence of thermal conduction and for Hildner's (dotted), Klimchuk-Raymond's (dashed), and CHIANTI-based (solid) loss functions. The shaded area denotes the region of instability at low temperatures obtained with the CHIANTI-based loss function that is not present for the other parametrizations. \label{fig:crit}}
\end{figure}

We perform a first-order expansion of  Equation~(\ref{eq:reldisper}) for  a low-$\beta$ plasma, where $\beta$ is the ratio of the gas pressure to the magnetic pressure. We obtain two different approximate dispersion relations, namely
\begin{equation}
\omega^2 - k^2 \Gamma^2 - \frac{\Lambda^2 \va^2}{\Gamma^2 - \Lambda^2} k^2 \sin^2 \theta \approx 0, \label{eq:appfast}
\end{equation}
for fast modes, and
\begin{equation}
\omega^2 - k^2 \Lambda^2 + \frac{\Lambda^2 \va^2}{\Gamma^2 - \Lambda^2} k^2 \sin^2 \theta \approx 0, \label{eq:appslow}
\end{equation}
for slow and thermal modes. Fast modes are weakly affected by nonadiabatic mechanisms and are not investigated further in the present work. We focus on Equation~(\ref{eq:appslow}) and explore in more detail the approximation for thermal modes.  Note that in the low-$\beta$ regime the propagation of slow and thermal modes is almost parallel to the magnetic field. By assuming $\Gamma^2 \gg \Lambda^2$, i.e, the Alfv\'en speed is much larger than the sound speed as typical for prominence conditions, Equation~(\ref{eq:appslow}) simplifies to 
\begin{equation}
\omega^2 - k^2 \Lambda^2 \left(  1 - \frac{\va^2}{\Gamma^2} \sin^2 \theta \right) \approx 0. \label{eq:appsol} 
\end{equation}
Now we use the definitions of $\Lambda^2$ and $\Gamma^2$ and expand Equation~(\ref{eq:appsol}) as a third order polynomial in $\omega$. Since the thermal mode is a purely imaginary solution we write $\omega = \omega_{\rm R} + i s$, where $s$ is the thermal mode growth rate, and set $\omega_{\rm R} = 0$. Because of the temporal dependence $\exp \left( - i \omega t \right)$, thermal mode perturbations are proportional to $\exp (st)$, meaning that for $s>0$ perturbations grow in time. This behavior corresponds to instability. On the contrary, for $s<0$ thermal disturbances are damped. To obtain an approximation for $s$, we neglect terms with $\mathcal O \left( s^2 \right)$ in the polynomial expansion of Equation~(\ref{eq:appsol}). The neglected terms are related to the slow mode and produce minor corrections to the thermal mode growth rate. After some algebraic manipulations, we obtain the approximate growth rate, namely
\begin{equation}
s \approx  - \frac{\left[ \left( \tilde{\kappa}_\perp \sin^2 \theta + \tilde{\kappa}_\parallel \cos^2 \theta \right) k^2 + \omega_{\rm T} - \omega_\rho \right] \va^2 \cos^2 \theta }{\left[ \left( \tilde{\kappa}_\perp \sin^2 \theta + \tilde{\kappa}_\parallel \cos^2 \theta \right) k^2 + \omega_{\rm T} - \omega_\rho \right] \eta_{\rm C} + \frac{\gamma \va^2 \cos^2 \theta}{\gamma - 1 }}  \label{eq:approxsfull}.
\end{equation}
In the absence of magnetic diffusion, $\eta_{\rm C} = 0$ and Equation~(\ref{eq:approxsfull}) becomes
\begin{equation}
s \approx - \frac{\gamma - 1}{\gamma} \left[ \left( \tilde{\kappa}_\perp \sin^2 \theta + \tilde{\kappa}_\parallel \cos^2 \theta \right) k^2 + \omega_{\rm T} - \omega_\rho \right]. \label{eq:approxs}
\end{equation}
Equation~(\ref{eq:approxs}) is similar to the expressions found by \citet{vanderlinden91} and \citet{solercont} for thermal continuum modes, and by \citet{carbonell09} and \citet{solerphd} for the imaginary part of the frequency of propagating thermal waves in a flowing medium. In the case without thermal conduction, i.e., $\kappa_\parallel = \kappa_\perp = 0$, the approximate growth rate is
\begin{equation}
s \approx - \frac{\gamma - 1}{\gamma} \left(  \omega_{\rm T} - \omega_\rho \right). \label{eq:approxs2}
\end{equation}
Equation~(\ref{eq:approxs2}) is independent of $k$, $\theta$, and the magnetic field strength and orientation.

  Equation~(\ref{eq:approxsfull}) provides us with the instability criterion. By taking into account the definitions of $\tilde{\kappa}_\parallel $, $\tilde{\kappa}_\perp$, $\omega_{\rm T} $ and $\omega_\rho$ (Equations~(\ref{eq:defs1}) and (\ref{eq:defs2})), we find that the combination of parameters to have $s>0$ in Equation~(\ref{eq:approxsfull}) must satisfy the condition
  \begin{eqnarray}
- \frac{\gamma - 1}{\gamma} \frac{\va^2 \cos^2 \theta}{\eta_{\rm C}} \frac{p}{T}  &<& \left( \kappa_\perp \sin^2 \theta + \kappa_\parallel \cos^2 \theta \right) k^2 \nonumber \\ &+& \rho \left( \frac{\partial L}{\partial T} \right)_\rho - \frac{\rho^2}{T} \left( \frac{\partial L}{\partial \rho} \right)_T < 0. \label{eq:crit}
  \end{eqnarray}
  To the best of our knowledge, Equation~(\ref{eq:crit}) is the first instability criterion for thermal modes that accounts for the effect of Cowling's diffusion. We deduce from Equation~(\ref{eq:crit}) that Cowling's diffusion has a stabilizing role since Cowling's diffusion incorporates a lower bound in the instability criterion.  In order to compare with previous instability criteria derived in the literature, we set $\eta_{\rm C} = 0$ and Equation~(\ref{eq:crit}) simplifies to
\begin{equation}
\left( \kappa_\perp \sin^2 \theta + \kappa_\parallel \cos^2 \theta \right) k^2 + \rho \left( \frac{\partial L}{\partial T} \right)_\rho - \frac{\rho^2}{T} \left( \frac{\partial L}{\partial \rho} \right)_T < 0. \label{eq:crit1}
\end{equation}
Equation~(\ref{eq:crit1}) agrees with the instability criterion provided by \citet{field} in his Equation~(25a) if the term accounting for perpendicular thermal conduction in added to Field's expression. For simplicity, we take the case $\eta_{\rm C} = 0$ and use the expression of the heat-loss function $L$ (Equation~(\ref{eq:heatloss})) to rewrite Equation~(\ref{eq:crit1}) in terms of parameters $\chi^*$ and $\alpha$. Then,  instability is present for values of $\alpha$ satisfying
\begin{equation}
 \alpha < 1 - \frac{ \left( \kappa_\perp \sin^2 \theta + \kappa_\parallel \cos^2 \theta \right) k^2}{\rho^2 \chi^* T^{\alpha -1}}. \label{eq:crit2}
\end{equation}
In the absence of thermal conduction ($\kappa_\parallel = \kappa_\perp =  0$) or for long wavelengths ($k \to 0$), the second term on the right-hand side of Equation~(\ref{eq:crit2}) vanishes and the instability criterion reduces to $\alpha < 1$.  Taking into account the values of the parameter $\alpha$ given in Table~\ref{tab:parameters}, the lowest  thermally unstable temperatures according to this criterion are $T \approx 8\times 10^4$~K in Hildner's fit, $T \approx 9.33 \times 10^4$~K in Klimchuk-Raymond's fit, and $T \approx 1.58\times 10^4$~K in the CHIANTI-based fit.  Importantly, we find that the threshold temperature for the thermal instability is substantially reduced using the CHIANTI-based radiative losses in comparison to Hildner's and Klimchuk-Raymond's functions.

Let us do a simple computation in the case without thermal conduction ($\kappa_\parallel = \kappa_\perp = 0$) and without Cowling's diffusion ($\eta_{\rm C} = 0$). In this case the approximate growth rate is given by Equation~(\ref{eq:approxs2}). Figure~\ref{fig:crit}  displays the growth rate computed from Equation~(\ref{eq:approxs2}) as a function of the temperature. A constant gas pressure of 6.64~mPa is assumed and the density is computed accordingly. As predicted by the instability criterion, there is a region of instability at low temperatures obtained with the CHIANTI-based loss function that is not present for the other parametrizations (see the shaded area in Figure~\ref{fig:crit}). This region of instability is present for temperatures between $1.58\times 10^4$~K and $3.16\times 10^4$~K. At these temperatures the prominence plasma is only partially ionized \citep[see, e.g.,][]{labrosse} and the roles of thermal conduction by neutrals and Cowling's diffusion may be relevant. These effects are investigated numerically in the Section~\ref{sec:res}.

Additionally, in Figure~\ref{fig:crit} we notice the abrupt jumps of the growth rate at the boundaries of the regions where different values of $\chi^*$ and $\alpha$ are used. The reason of these jumps is that, although the cooling function $L \left( \rho , T  \right)$ is continuous, its the derivatives  with respect to density and temperature are discontinuous where the values of parameters $\chi^*$ and $\alpha$ change.  Additional comments on this issue are given in \citet{vanderlindencont} and \citet{solercont}.

\begin{figure*}[!t]
\centering
  \includegraphics[width=0.95\columnwidth]{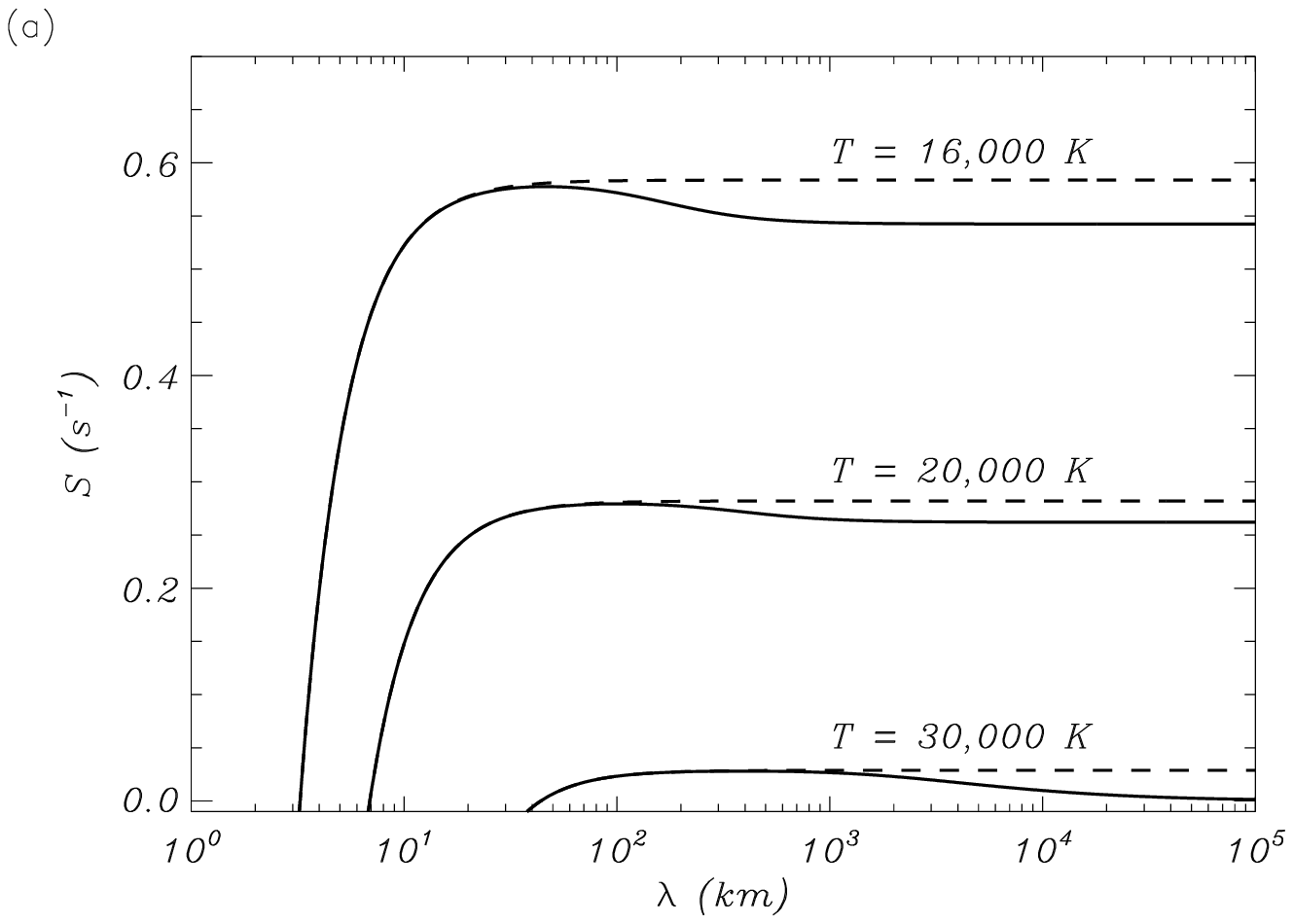}
    \includegraphics[width=0.95\columnwidth]{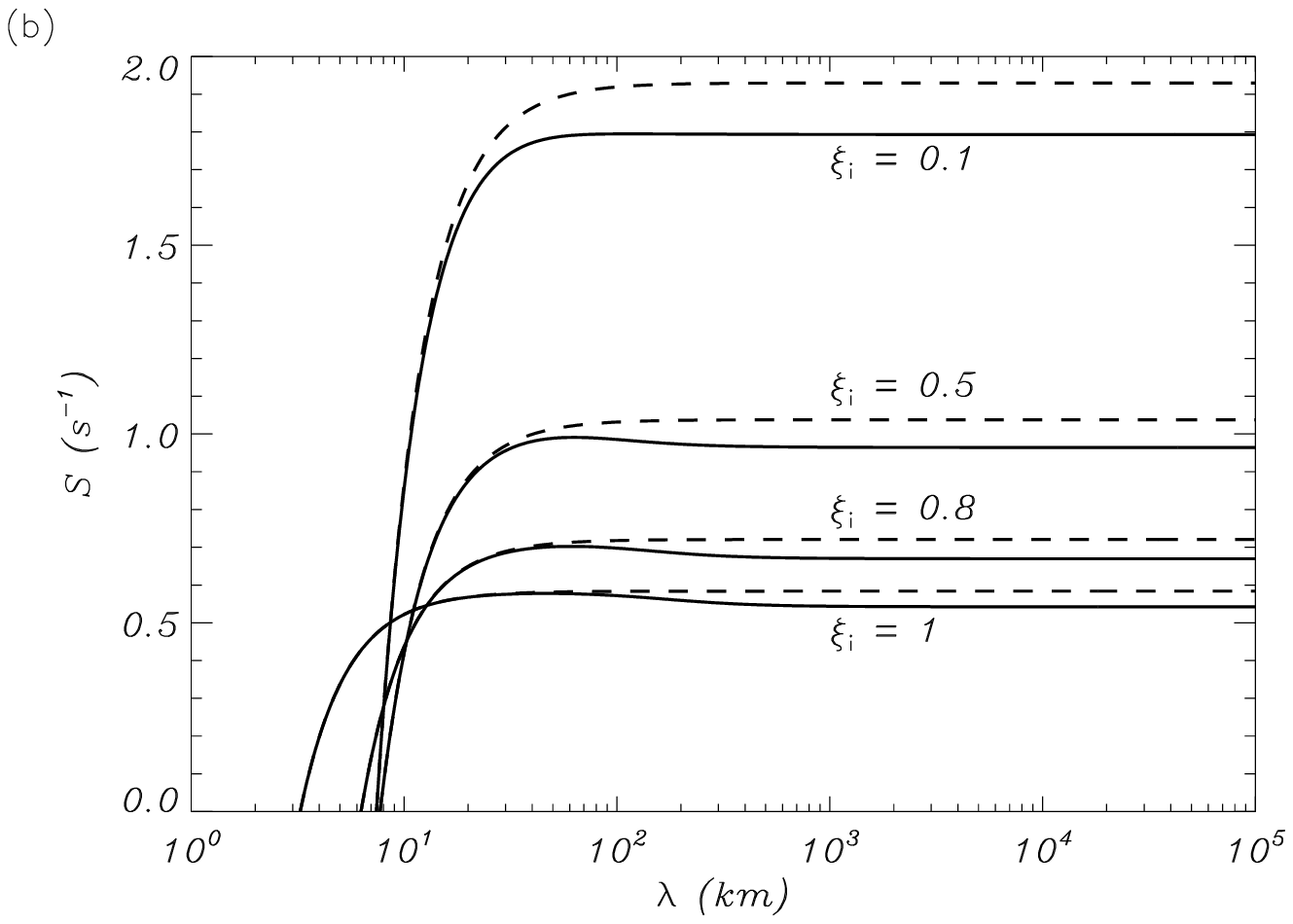}
\caption{(a)  Numerically computed growth rate (solid) and approximate value obtained from Equation~(\ref{eq:approxs}) (dashed) vs. wavelength for the CHIANTI-based loss function and different values of the temperature in the case of a fully ionized plasma, i.e., $\xi_{\rm i} = 1$. (b) Same as panel (a) but for $T=16,000$~K and different values of $\xi_{\rm i}$. \label{fig:growth}}
\end{figure*}

\section{Numerical results}
\label{sec:res}

Here we compute the thermal mode growth rate by solving the general dispersion relation (Equation~(\ref{eq:reldisper})) by standard numerical methods. We focus on the region of instability at low temperatures obtained with the CHIANTI-based loss function (shaded area in Figure~\ref{fig:crit}). In the following computations we use $B=10$~G, $p=6.64$~mPa, and $\theta = \pi/4$. These values of the equilibrium magnetic field strength and pressure are typical prominence parameters \citep[see, e.g.,][]{labrosserev}.  We compute the growth rate, $s$, as a function of the perturbation wavelength, $\lambda= 2 \pi / k$. Wavelengths typically observed in prominences are roughly between $10^3$~km and $10^5$~km  \citep[see][]{oliverballester02}.  This range of wavelengths correspond to disturbances usually detected from Doppler time series. We do not claim that all the observed disturbances are thermal modes. It is very difficult to determine the nature of the waves in the absence of additional information as, e.g., the velocity polarization or the magnetic and density perturbations. In particular, a clear distinction between slow and thermal modes may be very complicate \citep[see a discussion on this issue in][]{carbonell09}.  In this paper we use the observed wavelengths given by \citet{oliverballester02} as the most probable range of wavelengths for thermal modes.

\subsection{Fully ionized plasma}

First, we investigate the fully ionized case and set $\xi_{\rm i} = 1$. Therefore, the effect of thermal conduction by neutrals is absent and Cowling's diffusion becomes classical Ohm's diffusion, i.e., $\eta_{\rm C} = \eta$. Figure~\ref{fig:growth}(a) displays $s$ versus $\lambda$. As we focus on the unstable behavior of the thermal mode, only positive values of $s$ are displayed in Figure~\ref{fig:growth}(a). We consider three different temperatures within the region of instability denoted by a shaded area in Figure~\ref{fig:crit}. We compare the numerically computed growth rate (solid lines) with the approximation of Equation~(\ref{eq:approxs}) in the absence of diffusion, i.e., $\eta = 0$ (dashed lines). The approximate result is in good agreement with the actual growth rate. Different test computations with and without Ohm's magnetic diffusion (not displayed here for simplicity) indicate that magnetic diffusion has almost no impact on the value of the growth rate. This means that we can safely neglect the effect of diffusion and use the approximate growth rate given in Equation~(\ref{eq:approxs}) for the case $\eta = 0$. 

 Figure~\ref{fig:growth}(a) also shows that the thermal mode is stabilized for short wavelengths. The stabilization is due to thermal conduction by electrons. Thermal conduction becomes important as the wavelength decreases, so that the growth rate is reduced and the instability is suppressed for short enough wavelengths.  For long wavelengths the growth rate saturates and becomes independent of $\lambda$.  

Regarding the temperature, we find that the growth rate decreases as the temperature increases within the region of instability. For $T = 16,000$~K the maximum growth rate is $s\approx 0.6$~s$^{-1}$, while for $T = 30,000$~K the maximum growth rate decreases to $s\approx 0.05$~s$^{-1}$. These values of the linear growth rate indicate that the thermal instability operates in short timescales and suggest that the effect of the instability in prominences may be observable on the dynamics and evolution of cool plasma condensations. Nonlinear studies beyond the present linear analysis are needed in order to asses the actual impact of the instability on the condensation dynamics.

\subsection{Partially ionized plasma}

Here we incorporate the effects of partial ionization, namely thermal conduction by neutrals and Cowling's diffusion. In these computations we fix the temperature to $T=16,000$~K. Figure~\ref{fig:growth}(b) displays $s$ versus $\lambda$ for different values of $\xi_{\rm i}$. First of all, we obtain that when $\xi_{\rm i}$ decreases, the critical wavelength for stabilization increases due to thermal conduction by neutrals. At cool prominence temperatures, thermal conduction by neutrals is more efficient than conduction by electrons. Thus, the critical wavelength for stabilization is now determined by the conductivity of neutrals. We also see in Figure~\ref{fig:growth}(b) that the growth rate increases when $\xi_{\rm i}$ is reduced. The growth rate for $\xi_{\rm i} = 0.1$ is about four times larger, approximately, than that for  $\xi_{\rm i} = 1$. We can explain these result by taking into account that a constant gas pressure is assumed in the computations. So, when $\xi_{\rm i}$ decreases, the effective plasma density grows because of the increase of the amount of neutrals. The increase of the growth rate is a consequence of the increase of the effective density.

Also, we compare the numerical results (solid lines in Figure~\ref{fig:growth}(b)) with the approximate growth rates (dashed lines) given by Equation~(\ref{eq:approxs}). Note that Equation~(\ref{eq:approxs}) misses the effect of Cowling's diffusion. Nevertheless a reasonably good agreement between both results is obtained. The differences get larger when  $\xi_{\rm i}$ becomes small. As happens for Ohm's diffusion, we find that Cowling's diffusion have little influence on the thermal mode growth rate. Although Cowling's diffusion imposes a  lower bound for the instability criterion (see Equation~(\ref{eq:crit})),  its influence on the growth rate is of almost no relevance for realistic values of $\eta_{\rm C}$.

\section{Discussion}
\label{sec:discussion}

In this paper we have investigated the stability of thermal modes in partially ionized prominence plasmas in the single-fluid approximation. We have restricted ourselves to the linear phase and have derived an instability criterion that takes into account the effects of thermal conduction by electrons and neutrals, and Cowling's diffusion. We have applied the instability criterion using Hildner's \citep{hildner} and Klimchuk-Raymond's \citep{klimchuk} loss functions, which are frequently used in the literature, along with a new loss function derived from the CHIANTI atomic database \citep{parenti2006,parentivial}. Results using both Hildner's and Klimchuk-Raymond's loss functions predict the presence of thermal instability for temperatures higher than $10^5$~K, approximately. However,  the threshold temperature is significantly reduced for about an order of magnitude when the more up-to-date function based on the CHIANTI database is used.  In particular, instability can happen at temperatures as low as 15,000~K, approximately. Effects due to partial ionization, specially thermal conduction by neutrals, are relevant at these low  temperatures.

Focusing on the region of instability at low temperatures obtained with the new CHIANTI-based loss function, we have performed a parametric study of the linear growth rate by numerically solving the general dispersion relation. For constant gas pressure, we find that the growth rate decreases as the temperature increases. In addition, the growth rate increases as the amount of neutrals gets larger. We also find that thermal conduction reduces the growth rate for short wavelengths, with conduction by neutrals being more efficient than conduction by electrons in partially ionized plasmas. However, neither Ohm's nor Cowling's diffusion have an important influence on the growth rate.

A few remarks should be made about the applicability of our results. We must note that assumption of optically thin plasma might no be valid at low prominence temperatures. Due to finite optical thickness, the actual radiative losses of the plasma would be reduced in comparison to the optically thin case, so that the thermal mode growth rate would decrease consistently \citep[see][]{carbonell06}. Some attempts to incorporate the effect of finite optical thickness in Hildner's parametrization can be found in, e.g., \citet{rosner,milne}. To our knowledge, the effect of  finite optical thickness has not been incorporated in more up-to-date loss functions. We  recall that  the CHIANTI database is still incomplete at temperatures of the order of $T\sim10^4$~K. Further increase of the loss function at low temperatures is expected for more complete calculations. Thus, these two effects, namely the decrease of the radiative losses due to finite optical thickness and the increase due to the incorporation of additional line emissions, would determine the actual shape of the loss function at low temperatures and, therefore, the actual value of the growth rate of the unstable thermal modes. Also, the growth rates obtained here may be compared to the characteristic time scale for atomic processes. This means that the assumptions of ionization equilibrium of the plasma and electron thermal distribution might not be strictly valid at these short time scales. We plan to further investigate this aspect 
in the near future.

We conclude that thermal instabilities may take place in prominences at lower temperatures than those  predicted with previously existing loss functions. This may be important for the dynamics and energy balance of the prominence plasma. The obtained growth rates suggest that this low-temperature instability may have an observable effect in prominences.  For example, this low-temperature instability may help to form density enhancements in regions where the plasma is already cool as in, e.g., the prominence threads observed in H$\alpha$ images. However, nonlinear studies beyond the present normal mode approach are needed to assess the actual impact of the instability on the evolution of the prominence medium. This is relegated to future works.

\acknowledgements{
 The authors acknowledge discussion within ISSI Team on Solar Prominence Formation and Equilibrium: New data, new models. RS acknowledges support from a Marie Curie Intra-European Fellowship within the European Commission 7th Framework Program  (PIEF-GA-2010-274716). RS and JLB acknowledge the financial support received from the Spanish MICINN under project AYA2011-22846. RS and JLB also thank the financial support  from CAIB through the ``Grups Competitius'' scheme. SP acknowledges the support from the Belgian Federal Science Policy Office through the ESA-PRODEX programme. CHIANTI is a collaborative project involving George Mason University, the University of Michigan (USA), and the University of Cambridge (UK).}

\end{document}